\documentclass[sn-aps,iicol,pdflatex]{sn-jnl}% American Physical Society (APS) Reference Style
\usepackage{graphicx}%
\usepackage{multirow}%
\usepackage{amsmath,amssymb,amsfonts}%
\usepackage{amsthm}%
\usepackage{mathrsfs}%
\usepackage[title]{appendix}%
\usepackage{xcolor}%
\usepackage{textcomp}%
\usepackage{manyfoot}%
\usepackage{booktabs}%
\usepackage{algorithm}%
\usepackage{algorithmicx}%
\usepackage{algpseudocode}%
\usepackage{listings}%
\usepackage{graphicx}%
\usepackage{multirow}%
\usepackage{amsmath,amssymb,amsfonts}%
\usepackage{amsthm}%
\usepackage{mathrsfs}%
\usepackage[title]{appendix}%
\usepackage{xcolor}%
\usepackage{textcomp}%
\usepackage{manyfoot}%
\usepackage{booktabs}%
\usepackage{algorithm}%
\usepackage{algorithmicx}%
\usepackage{algpseudocode}%
\usepackage{listings}%
\usepackage{bm,physics}
\usepackage{amsmath,amssymb}
\usepackage{ulem, color}
\usepackage{amsmath}
\usepackage{natbib}
\bmdefine{\ba}{a}
\bmdefine{\bb}{b}
\bmdefine{\bx}{x}
\bmdefine{\by}{y}
\bmdefine{\bz}{z}
\bmdefine{\bn}{n}
\bmdefine{\bp}{p}
\newcommand{\BM}{\begin{pmatrix}}
\newcommand{\EM}{\end{pmatrix}}
\renewcommand{\d}{\dagger}

\newcommand{\hphi}{\hat\varphi}
\newcommand{\hpsi}{\hat\psi}

\newcommand{\intx}{\int\! d^3x\;}
\newcommand{\intxd}{\int\! d^3x'\;}
\newcommand{\intxxd}{\int\! d^3x\,d^3x'\;}

\newcommand{\ex}{\mathrm{ex}}
\begin{document}

%\title[Article Title]{Article Title}
\title {Bose-Einstein condensation of five $\alpha$ clusters in $^{20}$Ne and 
 supersolidity  
}
\author*[1]{\fnm{S.} \sur{Ohkubo}}\email{ohkubo@rcnp.osaka-u.ac.jp}

\author[2]{\fnm{J.} \sur{Takahashi}}%\email{d.junichi.takahashi@gmail.com}

\author[3]{\fnm{Y.} \sur{Yamanaka}}%\email{yamanaka@waseda.jp}

\affil*[1]{\orgdiv{Research Center for Nuclear Physics}, \orgname{Osaka University}, \orgaddress{\street{} \city{Ibaraki}, \postcode{567-0047}, \state{} \country{ Japan}}}

\affil[2]{\orgdiv{Faculty of Economics}, \orgname{Asia University}, \orgaddress{\street{} \city{ Tokyo}, \postcode{ 180-0022}, \state{} \country{Japan}}}

\affil[3]{\orgdiv{Institute of Condensed-Matter Science}, \orgname{Waseda University}, \orgaddress{\street{} \city{ Tokyo}, \postcode{ 169-8555}, \state{} \country{Japan}}}

\abstract{We show that  the five $\alpha$ cluster states recently  observed in $^{20}$Ne, slightly above the five $\alpha$ threshold energy, are Bose-Einstein condensates
 of five $\alpha$ clusters. The states are described well using a superfluid cluster model, where the order parameter is defined. We suggest that  the   the five $\alpha$ states are fragmented. Theory predicts the emergence of a five $\alpha$ 
  rotational roton  band characterized by a large moment of inertia. This band is formed through roton excitations of the five $\alpha$ BEC vacuum and possesses dual properties of superfluidity and crystallinity, a property of  supersolidity. 
The persistent existence of such a roton  band 
  is discussed and  confirmed 
for the four $\alpha$  condensate above the four $\alpha$ threshold in $^{16}$O and the  three $\alpha$  condensate  above the  three $\alpha$ threshold in $^{12}$C.
}




\maketitle

\section{Introduction}\label{sec:1}
\par
In analogy to BCS theory in  nuclei \cite{Bohr1969B,Ring1980,Brink2005} collective motions of superfluidity caused by $\alpha$ cluster condensation
 have  received  attention within the  framework of  many-body theory over the last decades \cite{Eichler1972,Gambhir1983}.
Inspired by the observation of Bose-Einstein condensation (BEC) of trapped cold atoms \cite{Cornel2002},  extensive theoretical \cite{Tohsaki2001,Matsumura2004,Funaki2015,Funaki2016,Nakamura2016,Katsuragi2018} and experimental
\cite{Itoh2011A,Itoh2011B,Freer2009,Zimmerman2011,Zimmerman2013,Itoh2013,Freer2011} studies have focused on the BEC of  three $\alpha$ clusters in the Hoyle state and excited states with a well-developed $\alpha$ cluster structure above the $\alpha$ threshold energy in $^{12}$C. 
Also  the BEC of the four $\alpha$ clusters in $^{16}$O above the four $\alpha$ threshold  has been studied theoretically \cite{Tohsaki2001,Funaki2008C,Ohkubo2010,Funaki2018,Takahashi2020} and experimentally \cite{Chevallier1967,Freer1995,Freer2004,Freer2005,Itoh2014,Curtis2016,Barbui2018}.

\par
For the confirmation of   the BEC of $\alpha$ cluster structures, regardless of the number of $\alpha$ clusters in the lowest $0$s state ($\sim$70$\%$ for the Hoyle state in $^{12}$C \cite{Matsumura2004} and 61$\%$ for the $0^+$ state at 15.1 MeV in $^{16}$O \cite{Funaki2008C}), it is essential to observe the Nambu-Goldstone (NG) mode  due to global phase locking and the related collective motions of the BEC. The observation of  typical phenomena such as superfluidity \cite{Brink2005}, roton excitations \cite{Landau1941,Landau1947,Feynman1953A,Feynman1953B,Feynman1954,Feynman1957}, quantum vortex \cite{Cornel2002}, or the Josephson effect \cite{Josephson1966,Broglia2022}  would definitively confirm the existence  of BEC of $\alpha$ cluster structures.
A superfluid cluster model (SCM) based on effective field theory, in which the order parameter of BEC is defined and the NG mode operators originating in spontaneous symmetry breaking (SSB) of the global phase are properly formulated \cite{Nakamura2014}, has successfully confirmed the BEC of $\alpha$ cluster structure in $^{12}$C \cite{Nakamura2016,Katsuragi2018} and $^{16}$O\cite{Takahashi2020}.

\par
In He II, collective excitations involving a phonon of the NG mode and a roton
\cite{Landau1941,Landau1947,Feynman1953A,Feynman1953B,Feynman1954,Feynman1957} have been extensively  investigated \cite{Cowley1971,Woods1973,Andersen1994,Griffin1993,Glyde1995,%
Andersen1994B,Gibbs1999,Milner2023}.
These excitations were confirmed experimentally in inelastic neutron scattering \cite{Palevsky1957} and  the rotation of He II \cite{Vinen1958}.
In  dipolar Bose-Einstein condensates of  cold atoms \cite{Dell2003,Santos2003,Blakie2012,Bisset2013},  observations of roton excitations 
  suggest the   possibility of accessing a superfluid with order, a supersolid \cite{Chomaz2018,Petter2019,Schmidt2021}. 
The existence of a supersolid  in nature and the mechanisms behind its creation  \cite{Andreev1969,Chester1970,Leggett1970,Matsuda1970} have been long-standing questions since the discovery of He II  \cite{Kapitsa1938,Allen1938}.
Notably, recent reports include the observation of a supersolid in optically trapped cold atoms \cite{Tanzi2019,Bottcher2019,Chomaz2019,Biagioni2024} and the supersolidity of $\alpha$ cluster structures at low excitation energies in nuclei \cite{Ohkubo2020B,Ohkubo2022}.

\par
Recently Adachi { et al.} \cite{Adachi2020} observed  three five $\alpha$ cluster condensate candidate (0$ ^+$) states in $^{20}$Ne at $E_x$=21.2, 21.8 and 23.6 MeV in addition to the 22.5 MeV state observed by Swartz { et al.} \cite{Swartz2015}.
A five $\alpha$ cluster model calculation \cite{Zhou2023} yielded only
two candidates $0^+$ states. It is crucial to determine whether the observed four five $\alpha$ states in $^{20}$Ne are indeed BEC states by confirming the NG mode and the collective motions  associated with BEC. Also, understanding  the BEC structure in $^{20}$Ne and the four $\alpha$ (three $\alpha$) BEC structure in $^{16}$O ($^{12}$C), where the bands with the $\alpha$ cluster structure  are built on the NG mode $0^+$ state, in a unified way is essential.

\par
In this paper we show that the observed $\alpha$ cluster  states in $^{20}$Ne are BEC states by  using  the SCM and  predict the emergence of a rotational   band with a large moment of inertia.  The band  is shown to be caused by roton excitations  of the five $\alpha$ BEC vacuum and has a dual property of superfluidity and crystallinity, a property of a supersolid. 
The persistence of such a duality is discussed and  confirmed for the bands with the four $\alpha$ cluster condensate states in $^{16}$O and the  three $\alpha$ cluster condensate states in $^{12}$C.

\begin{figure}[t!]
\begin{center}
\includegraphics[width=8.6cm]{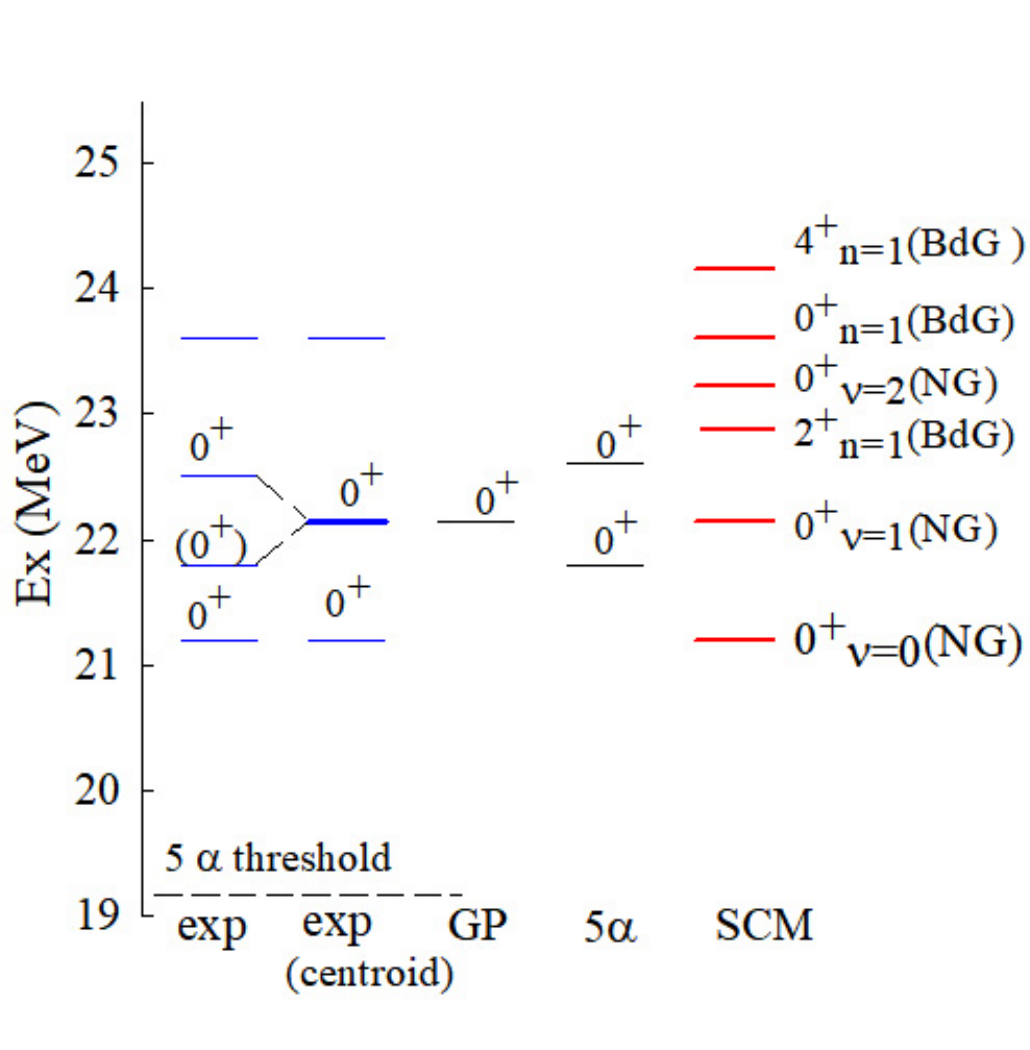}
\caption{  The experimental energy levels of the five $\alpha$ condensate candidate states in $^{20}$Ne \cite{Adachi2020,Swartz2015} are compared with the
 calculations, the
   SCM, the Gross-Pitaevskii model (GP) \cite{Yamada2004}, and the five $\alpha$ cluster model (5$\alpha$) \cite{Zhou2023}.  Fragmentation is assumed   for the  two levels from the centroid  indicated by the bold line. 
   }
\label{fig:fig1}
\end{center}
\end{figure}

\section{Five $\alpha$  condensation in $^{20}$Ne}\label{sec:2}
\par
In the SCM~\cite{Nakamura2014,Nakamura2016}, the model Hamiltonian for a bosonic field $\hpsi(x)$ $(x=(\bx,t))$ representing the $\alpha$ cluster is given as follows:
\begin{align}
&
\hat{H}=\intx \hpsi^\d(x) \left(-\frac{\nabla^2}{2m}+
V_\ex(\bx)- \mu \right) \hpsi(x) \notag\\
&
\,\,+\frac12 \intxxd \hpsi^\d(x)
\hpsi^\d(x') U(|\bx-\bx'|) \hpsi(x') \hpsi(x) \,.
\label{Hamiltonian}
\end{align}
The trap potential $V_\ex$ has 
$V_\ex(r)= m \Omega^2 r^2/2\,,$
and the $\alpha$--$\alpha$ interaction is given by
$ U (|\bm x -\bm x'|) = V_r e^{-\mu_r^2 |\bm x -\bm x'|^2}
- V_a e^{-\mu_a^2 |\bm x -\bm x'|^2}\,$ \cite{Ali1966}.
When BEC of $\alpha$ clusters occurs,
we decompose $\hpsi$
as $\hpsi(x)=\xi(r)+\hphi(x)$.
The Gross--Pitaevskii (GP) equation determines the order parameter $\xi(r)$
 by
$\left\{ -\frac{\nabla^2}{2m}+V_\ex(r) -\mu + V_H(r)
\right\} \xi(r) = 0 $\,
where
$ V_H(r) = \intxd U(|\bx-\bx'|)\xi^2(r')$
and $\mu$ is the chemical potential. 
$\xi$ is normalized
with the condensed particle number $N_0$ as
$\intx |\xi(r)|^2 = N_0\,.$
The collective excitations on the condensate are described by 
the Bogoliubov-de Gennes (BdG) equation.
The NG equation,
$\hat H_u^{QP} \ket{\Psi_\nu} = E_\nu \ket{\Psi_\nu}
(\nu=0,1,\cdots)\,$ 
 (see Refs.~\cite{Nakamura2014,Nakamura2016} for details),
determines the  NG mode $0^+$  states.
The excitation spectrum is obtained by solving these  three coupled equations.

\par
%we adjust 
The two parameters    $\Omega$ and $V_r$, which control the size and stability of the trapped condensate, respectively,  are determined  as in Refs.~\cite{Nakamura2016,Katsuragi2018,Takahashi2020}. 
The chemical potential is fixed by the input $N_0$=5.
Assuming a condensation rate of 50\%, considering $\sim$60\% of the four $\alpha$ condensate      $0^+$ state at 15.1 MeV in $^{16}$O \cite{Funaki2008C,Takahashi2020},
     we initially considered the lowest excited state at 21.2 MeV above the five $\alpha$  threshold as the vacuum. 
  However, although the SCM reproduces all the observed states, 
the resulting calculated rms radius $\overline{r}$ of the 21.2 MeV $0^+$ state (4.29 fm) is significantly smaller than the $\overline{r}$=5.62 fm of the calculated four $\alpha$ condensate $0^+$ state at 15.1 MeV in $^{16}$O \cite{Takahashi2020}, rendering it physically unacceptable.  This discrepancy suggests that at least one of the observed four states is fragmented. 

\par
In Fig.~\ref{fig:fig1} the energy levels calculated with $\Omega$=1.06 MeV/$\hbar$ and $V_r$= 505 MeV are displayed. These calculations are based on taking the 21.2 MeV $0^+$ state as the vacuum and assuming that the two states at 21.8 and 22.5 MeV are fragmented from the centroid. 
The resulting  calculated  vacuum radius $\overline{r}$=6.04 fm 
is reasonable,  comparable to the 5.5 fm  in  Ref.~\cite{Yamada2004}.
 The calculation reproduces the experimental (centroid) data and locates the BdG $2^+$ state at 22.9 MeV and $4^+$ state  at 24.2 MeV.
The  NG mode $0^+$ states   appear at 22.2 and   23.2 MeV.
On the other hand,  Refs.~\cite{Yamada2004,Zhou2023} only  calculated the $0^+$ states,
which are   slightly  higher than   the  experimental data.

\par
\begin{figure}[t!]
\begin{center}
\includegraphics[width=8.6cm]{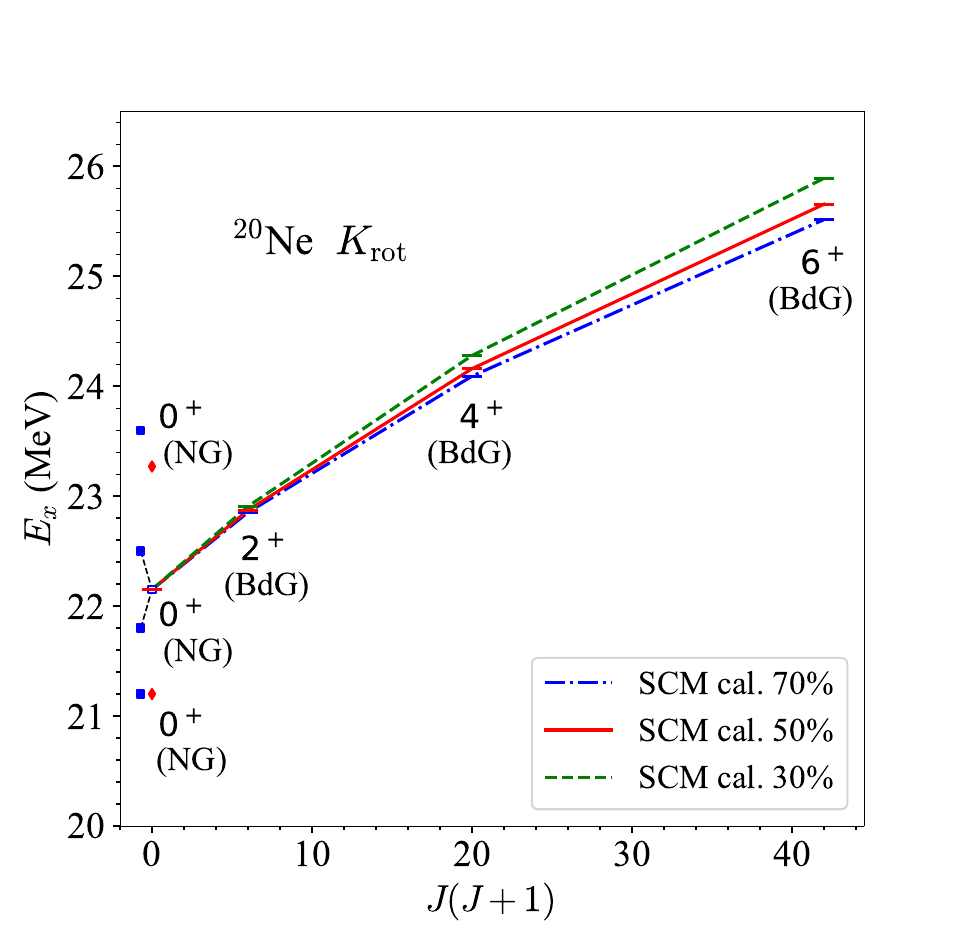}
\caption{ The  calculated rotational band,  $K_{\rm rot}$,  excited by rotons is built on the NG mode $0^+$ of the five $\alpha$ BEC in $^{20}$Ne for the condensation rates 30\%, 50\% and 70\%. 
The experimental data are indicated by blue squares,  and the lines are to guide the eye.
}
\label{fig:band}
\end{center}
\end{figure}

 In Fig.~\ref{fig:band}, it is surprising that a rotational band, $K_{\rm rot}$, emerges on the NG mode $0^+$ state.
In the SCM, we assume no geometrical intrinsic configuration for the five $\alpha$ clusters trapped in the spherical harmonic oscillator potential. 
Although it is not self-evident that the states form a rotational band from the viewpoint of field theory, the emergence of the $K_{\rm rot}$ band
indicates that it has a common geometrically deformed intrinsic configuration for the five $\alpha$ clusters. 
While the homogeneous He II system has a continuous spectrum, which consists of phonon excitations without angular momentum and roton ones with angular momentum, separated by the gap $\Delta$ \cite{Landau1941,Landau1947,Feynman1953A,Feynman1953B,Feynman1954,Feynman1957}, 
the $\alpha$ cluster system of the SCM is inhomogeneous due to the trap, resulting in discrete NG and BdG levels. We may identify the NG and BdG states in the SCM with the phonon and roton excitations in He II, respectively. 
Consequently, the states of the $K_{\rm rot}$ band correspond to roton excitations, and the energy of its head provides the gap $\Delta$.
As indicated in Fig.~\ref{fig:band}, the emergence of the roton rotational band remains robust for different condensation rates of 30\%, 50\%, and 70\%, with little change in the moment of inertia.
Our calculations also confirm the appearance of the rotational roton band $K_{\rm rot}$ when the lowest   two or three states are assumed to be fragmented from each centroid energy.

 %%Table monopole BE2 Table I
 \begin{table}[b]
\caption{ $B({\rm E}2)$ values of the  $E2$ transitions in $^{20}$Ne calculated in the SCM  for the condensation rates, (a) 30\%, (b) 50\%   and (c) 70\%, in units of $e^2\,{\rm fm}^4$. }
\label{tab:BE2}
%	\begin{center}
		\begin{tabular}{lrrr} \hline\hline
		\multicolumn{1}{c}{Transition}  	        		& (a)&  (b) & (c) 
		   	        		\\ \hline
		   	       $B({\rm E}2:2^+_{n=1}({\rm BdG})\, \rightarrow \, 0^+_{\nu=0}({\rm NG}))$ &440& 786 &1150\\
			$B({\rm E}2:2^+_{n=1}({\rm BdG})\, \rightarrow \, 0^+_{\nu=1}({\rm NG}))$ &1690 &2182  &2541\\
			$B({\rm E}2:4^+_{n=1}({\rm BdG})\, \rightarrow \, 2^+_{n=1}({\rm BdG}))$ &1283 &1487 	&1628\\
			$B({\rm E}2:6^+_{n=1}({\rm BdG})\, \rightarrow \, 4^+_{n=1}({\rm BdG}))$ &1761& 2070 &2301\\
 \hline \hline
\end{tabular}
%	\end{center}
\end{table}

\begin{figure}[t!]
\begin{center}
\includegraphics[width=7.6cm]{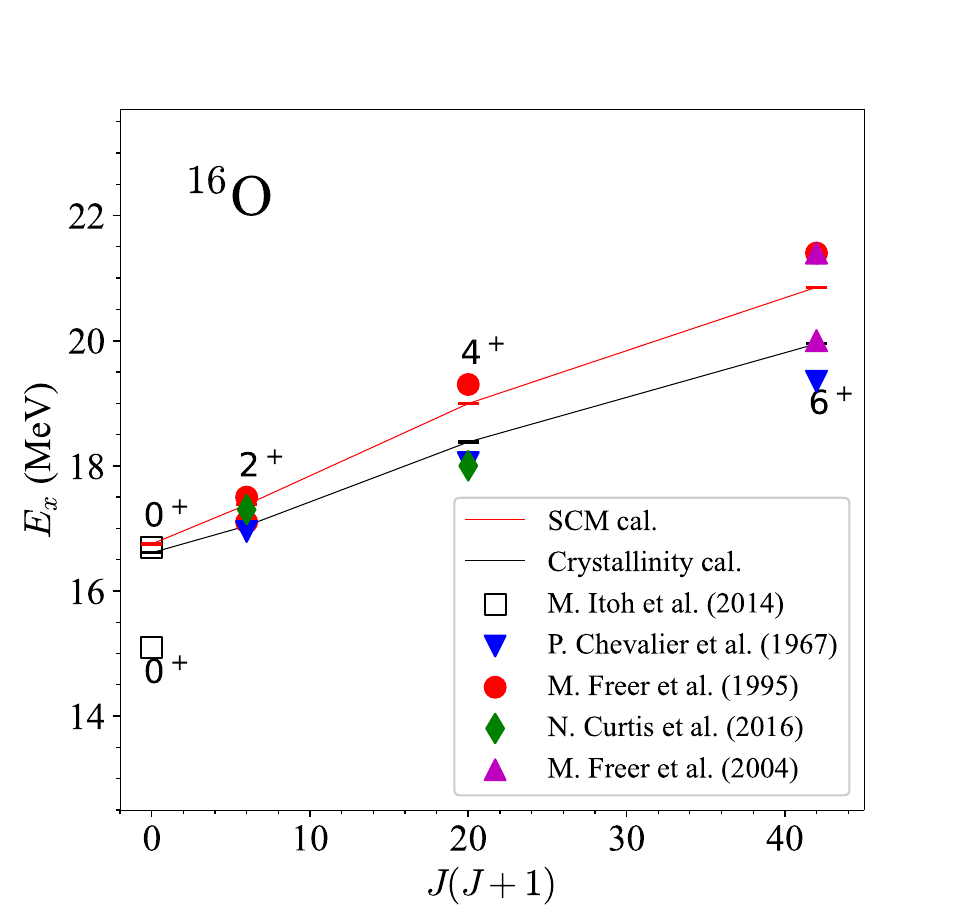}
\caption{ The four $\alpha$ cluster rotational  band $K_{\rm rot}$ in $^{16}$O. The theoretical calculations using the geometrical cluster model (crystallinity) \cite{Ohkubo2010}  and the SCM  \cite{Takahashi2020}  both  reproduce  the fragmented  experimental states of the  band   from Ref.~\cite{Freer2004}  and   Refs.~\cite{Chevallier1967,Freer2005,Curtis2016}, respectively.  The lines are to guide the eye. 
 }
\label{fig:16Oband}
\end{center}
\end{figure}

\par
The rotational constant, $k$=$\frac{\hbar^2}{2\mathcal{J}}$, is estimated to be 67 keV from the band shown in Fig.~\ref{fig:band}, where $\mathcal{J}$ represents the moment of inertia. Comparatively, $k=$147 keV for the $K=0_1^-$ parity-doublet band starting at $E_x$=5.79 MeV, while $k=$101 keV for the $K=0_4^+$ higher nodal band beginning at $E_x\approx$8.78 MeV with a very developed $\alpha$ cluster structure in $^{20}$Ne.
Notably, the $\alpha$ cluster band $K_{\rm rot}$ in Fig.~\ref{fig:band} has a significantly larger moment of inertia — about 3.0 times that of the $K=0^+_1$ ground band in $^{20}$Ne. Accordingly, Table~\ref{tab:BE2} predicts large $B(E2)$ values for the roton band states, and similarly large values are obtained for the condensation rates of 30\% and 70\%. We note that the $E2$ transition from the $2^+_{n=1}$(BdG) state to the band head $0^+_{\nu=1}$(NG) state is much stronger than that to the $0^+_{\nu=0}$(NG) state of the vacuum, reinforcing that the roton band is built on the first excited NG mode state.

 %Fig.4 robust roton band 40% 60%  80% 100% 16O
\begin{figure}[t!]
\begin{center}
\includegraphics[width=8.6cm]{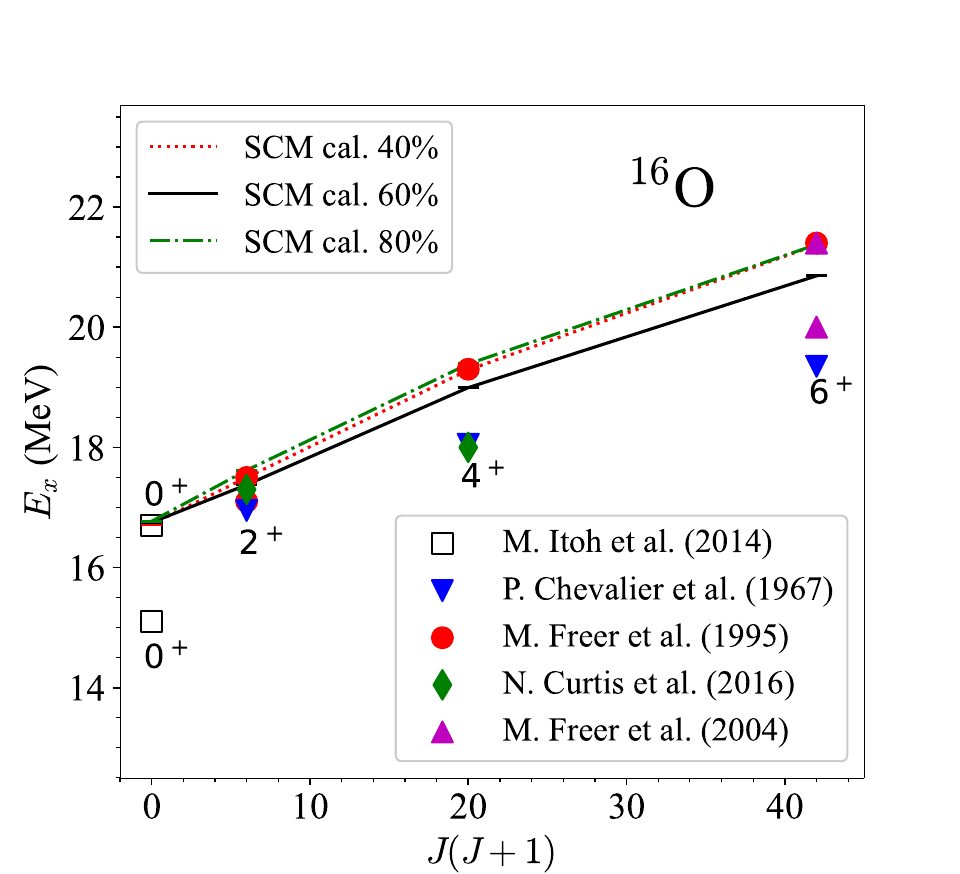}
\caption{The emergence of the   four $\alpha$ rotational  band $K_{\rm rot}$  in $^{16}$O in the SCM calculations \cite{Takahashi2020}  for the different condensation rates: 40, 60, and 80\%.  
The lines are to guide the eye. 
}
\label{fig:roton-stability}
\end{center}
\end{figure}
 
\par
The  field theoretical excitations described above can be explained from the viewpoint of crystallinity,  the geometrical configuration of the five $\alpha$ clusters. 
Since the vacuum $0^+$ state is spherical, the excited band states ($0^+$, $2^+$, $4^+$ and $6^+$) are attained by separating one $\alpha$ cluster with the relative orbital angular momentum $L=$0, 2, 4 and 6, respectively. 
The band head $0^+$ with $L=0$  represents a higher nodal $0^+$ state, in which the relative motion is excited to have one more node orthogonal to the vacuum. 
The $K_{\rm rot}$ band states form a rotational band with the configuration $\alpha$ + (core), where the core  consists of a four $\alpha$ condensate of $^{16}$O.
Thus, the $K_{\rm rot}$ band states  exhibit  not only superfluidity of roton excitations
but also a deformed geometrical structure, a duality of superfluidity and crystallinity. %\cite{Ohkubo2022}. 
A system with this duality is   a supersolid,  which has been long sought in He II \cite{Andreev1969,Chester1970,Leggett1970,Matsuda1970}. 
A promising experimental approach to confirm BEC of the five $\alpha$ clusters in $^{20}$Ne may  involve observing the predicted rotational roton  band due to roton excitations 
above the five $\alpha$ threshold, which is highly desired.

\section{Discussion}\label{sec:3}
\par
Since the rotational band  $K_{\rm rot}$ has not yet  been observed  in $^{20}$Ne, it is important to know whether such a band exists in $\alpha$ cluster nuclei where $\alpha$  condensation has been intensively investigated theoretically and experimentally. Such candidate nuclei include  $^{12}$C and $^{16}$O.

\par
First we discuss $^{16}$O. 
In Fig.~\ref{fig:16Oband} the observed four $\alpha$ cluster states \cite{Chevallier1967,Freer1995,Freer2004,Freer2005,Itoh2014,Curtis2016,Barbui2018},
which are fragmented, are displayed. These states form a rotational band with a very large moment of inertia,  have long been believed to have a geometrically linear chain of four $\alpha$ clusters, as  proposed in Ref.~\cite{Chevallier1967}. 
From the viewpoint of $\alpha$ condensation, it was shown that the band can be better  understood as  having  a geometrically  $\alpha$+$^{12}$C(g.s, $0_2^+$) configuration, rather than the linear chain, in   coupled channel cluster model calculations in Ref. \cite{Ohkubo2010}.

\par

On the other hand, the band states can also be well understood from the compelling viewpoint of a superfluid cluster model without assuming any geometrical configurations for four $\alpha$ clusters \cite{Takahashi2020}. As seen in Fig.~\ref{fig:16Oband}, both pictures reproduce the experimental band almost equally. The two compelling properties of the band have not been understood in the literature consistently and in a unified way.

We suggest, as noticed in $^{20}$Ne theoretically, that the band states have a dual property of crystallinity and superfluidity. The band states with angular momentum are understood as roton excitations of the condensate vacuum, i.e., the rotational band states $J=2^+$, $4^+$, and $6^+$ are understood to be caused by the roton excitations with angular momentum $J=2$, $4$, and $6$, from the condensate vacuum $0^+$ state at 15.1 MeV. The band head $0^+$ state is an NG state. To the authors' best knowledge, the roton excitations for finite systems like nuclei have never been noticed.

 %Fig5  12C roton band 
\begin{figure}[t!]
\begin{center}
\includegraphics[width=8.6cm]{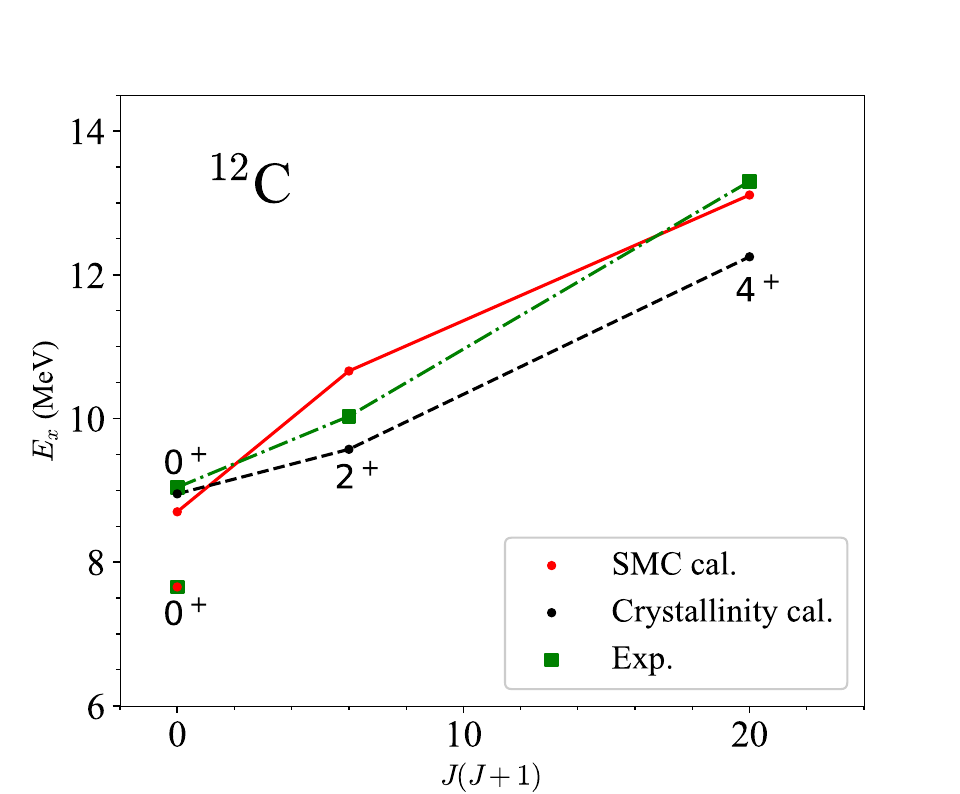}
\caption{The three $\alpha$ rotational  band $K_{\rm rot}$ in $^{12}$C. The SCM \cite{Nakamura2016}  and the  geometrical three $\alpha$ cluster model (crystallinity) \cite{Ohtsubo2013,Ohtsubo2014} reproduce
the experimental  band   from Refs.~\cite{Freer2011,Itoh2011A,Itoh2011B,Zimmerman2013}.
 The lines are to guide the eye. 
 }
\label{fig:12Cband}
\end{center}
\end{figure}

\par
The geometrical cluster model's  $\alpha$+$^{12}$C($0_2^+$) configuration  is consistently understood from the SCM calculation as follows:
In the crystallinity picture, the $0^+$ band head in Fig.~\ref{fig:16Oband} corresponds to a state where one of the four $\alpha$ clusters in the vacuum is lifted to the next higher $s$ orbit,  resulting in a nonspherical  $\alpha$+$^{12}$C$({0_2^+})$ structure. 
The relative wave function has  an additional node due to its orthogonality to the vacuum. Due to the SSB of  rotational invariance, a rotational band with the $\alpha$+$^{12}$C$({0_2^+})$ cluster structure emerges.
In other words, the $\alpha$ cluster rotates around the $^{12}$C($0_2^+$) core 
with  orbital angular momentum $L$=0, 2, 4, and 6, corresponding to
 phonon (for $L$=0) and roton (for $L\ne$0) excitations of the vacuum in the field theoretical picture. The rotational constant $k$=80 keV
calculated in the SCM  agrees well with  the estimated $k=$86 keV   from 
 the crystallinity picture  \cite{Ohkubo2010}.
The fact that both pictures reproduce the experimental data of the four  $\alpha$ condensate  in Fig.~\ref{fig:16Oband} indicates that  the band exhibits the duality of superfluidity and crystallinity \cite{Ohkubo2022}, characteristic of a supersolid. 
The large $\mathcal{J}$ is the consequence of  the roton excitations 
and  it may serve as a fingerprint of  supersolidity.
 This is understood intuitively because   $\mathcal{J}$ increases in  $J\hbar$=$\mathcal{J}\omega$  as the rotational frequency $\omega$ approaches  zero due to superfluidity.  

 In Fig.~\ref{fig:roton-stability}, one  observes that the appearance of the roton  band $K_{\rm rot}$  in $^{16}$O remains robust and almost unchanged for  different condensation rates as  for $^{20}$Ne shown in  Fig.~\ref{fig:band}.

\par
Next, we discuss $^{12}$C. Since it was demonstrated that the concept of the $\alpha$ cluster condensate, with the dual properties of crystallinity and superfluidity, persists in the experimental data of $^{16}$O, it is expected that this concept may also persist in the three $\alpha$ system of $^{12}$C, where most thorough study has been devoted both experimentally and theoretically.

As displayed in Fig.~\ref{fig:12Cband}, the experimental states with a well-developed three $\alpha$ cluster structure in $^{12}$C -- $0_3^+$ (9.04 MeV) \cite{Itoh2011A,Itoh2011B}, $2_2^+$ (10.03 MeV) \cite{Zimmerman2013}, and $4_1^+$ (13.3 MeV) \cite{Freer2011} -- form a rotational band. 
In the crystallinity picture, the band head $0_3^+$ is created by separating one $\alpha$ cluster from the spherical Hoyle state to a higher nodal state with $L=0$, resulting in the $\alpha$+$^{8}$Be structure. In fact, the $0_3^+$ state was shown to be the higher nodal state in geometrical three $\alpha$ cluster model calculations using the orthogonality condition model (OCM) \cite{Kurokawa2005,Kurokawa2007}. More precise geometrical three $\alpha$ cluster OCM calculations confirmed the $\alpha$+$^{8}$Be structure of the $0_3^+$ state \cite{Ohtsubo2013,Ohtsubo2014}. The existence of the $2^+$ and $4^+$ states built on the $0_3^+$ state, which was first predicted in Ref.~\cite{Kurokawa2005,Kurokawa2007}, was definitively confirmed in Ref.~\cite{Ohtsubo2013,Ohtsubo2014}. The calculation based on the crystallinity picture reproduces the experimental rotational band as seen in Fig.~\ref{fig:12Cband} (dashed line).

  On the other hand, 
 the band is  also well reproduced   in the SCM calculation, as indicated  by the solid line in  Fig.~\ref{fig:12Cband}. 
The calculation also predicts a $6^+$at 15.4 MeV in the roton band .
In the SCM,  the band head $0_3^+$ is understood to be an NG mode state, while  the $2^+$, $4^+$ and $6^+$ 
 band states   result from  roton excitations of the vacuum Hoyle $0^+$ (7.65 MeV) state. The results in  Fig.~\ref{fig:12Cband} are calculated  with   $\Omega$=1.80 MeV/$\hbar$, $V_r$=422 MeV and $N_0$=3, as in Ref.~ \cite{Nakamura2016}. 
We note that  $\alpha$ cluster model calculations  based on the $\alpha$ condensation picture of Ref.~\cite{Funaki2015} with no order parameter and no particle number fluctuations do not reproduce the experimental rotational band, locating the $2^+_2$ state lower than the band head $0_3^+$.

  Thus,  it is also found that the concept of duality 
     of superfluidity and  crystallinity,  persists in  the observed rotational band  with a large moment of inertia,  built just above the condensate vacuum, the Hoyle state. 
 It is challenging to observe a  roton band  with supersolidity  in other  nuclei such as $^{24}$Mg and  $^{28}$Si, which will serve  as a unique fingerprint of BEC of the $\alpha$ clusters.

\section{Summary}\label{sec:4}
\par
To summarize, we have shown that the recently observed five $\alpha$ cluster states above the five $\alpha$ threshold in $^{20}$Ne are BEC states, using a superfluid cluster model. The emergence of a rotational band due to roton excitations, the roton band, is predicted in $^{20}$Ne.

The existence of the rotational roton bands with $\alpha$ cluster structure of four $\alpha$ cluster states in $^{16}$O and three $\alpha$ cluster states in $^{12}$C is discussed and demonstrated in the available experimental data. The roton bands were shown to exhibit the dual properties of superfluidity and crystallinity—a characteristic of a supersolid.

The observation of a roton band, characterized by a large moment of inertia, may serve as a fingerprint for Bose-Einstein condensation of $\alpha$ clusters and supersolidity.
 It is highly desired to observe the predicted roton band states in $^{20}$Ne experimentally.

\bmhead{Acknowledgements}
The authors thank the Yukawa Institute for Theoretical Physics at Kyoto University for the hospitality extended during a stay in March 2024.

%\begin{itemize}
%No funding.
%\item Conflict of interest/Competing interests 
%Not applicable
%(check journal-specific guidelines for which heading to use)
%\item Ethics approval and consent to participate 
%The authors agree with this.
%\item Consent for publication
%The authors agree with publication of the manuscript in the traditional publishing model.
%\item Data availability 
%This manuscript has no associated data.

%This manuscript has no associated Materials.
%This manuscript has no associated code/software. 
%These authors contributed equally to this work.
%\end{itemize}

%\noindent
%If any of the sections are not relevant to your manuscript, please include the heading and write `Not applicable' for that section. 


\begin{thebibliography}{00}
\bibitem {Bohr1969B}
A. Bohr,  B. R. Mottelson: {\it Nuclear Structure}, Vol. II,
(Benjamin, Inc., New York, 1975).
\bibitem {Ring1980}
P. Ring, P. Schuck,
{\it The Nuclear Many-Body Problem} (Springer-Verlag, Berlin, 1980).
\bibitem{Brink2005}
D. M. Brink, R. A. Broglia,
{\it Nuclear Superfluidity: Pairing in Finite Systems} 
(Cambridge University Press, Cambridge, 2005).
\bibitem {Eichler1972}
J. Eichler,  M. Yamamura, 
Nucl. Phys.   A {\bf  182}, 33 (1972).
\bibitem {Gambhir1983}% A superfluid condensate of alpha particles 
Y. K. Gambhir, P. Ring,  P. Schuck, 
Phys. Rev. Lett. {\bf 51}, 1253 (1983).
\bibitem {Cornel2002}
E. A. Cornell,  C. E. Wieman, 
Rev. Mod. Phys. {\bf 74}, 875 (2002).
%Rev. Mod. Phys. {\bf 74}, 875 (2002); see also references therein.
%12C theory 3 alpha BEC
\bibitem {Tohsaki2001}
A. Tohsaki, H. Horiuchi, P. Schuck,  G. R$\ddot{\rm o}$pke,
Phys.~Rev.~Lett. {\bf 87}, 192501 (2001).
%12C Hoyle
\bibitem {Matsumura2004}
H.~Matsumura,  Y.~Suzuki,
Nucl.~Phys.~A {\bf 739}, 238 (2004).
\bibitem{Funaki2015} %Hoyle band 12C
Y. Funaki, Phys. Rev. C {\bf 92}, 021302(R) (2015).
\bibitem {Funaki2016} %12C Hoyle
Y. Funaki, 
Phys. Rev. C {\bf 94}, 024344 (2016).

\bibitem {Nakamura2016}
Y.~Nakamura, J.~Takahashi, Y.~Yamanaka,  S.~Ohkubo,
Phys.~Rev.~C {\bf 94}, 014314 (2016); {\bf 98}, 049901(E) (2018).
\bibitem {Katsuragi2018}
R.~Katsuragi, Y.~Kazama, J.~Takahashi, Y.~Nakamura, Y.~Yamanaka,  S.~Ohkubo,
Phys.~Rev.~C {\bf 98}, 044303 (2018).
\bibitem {Freer2009} % observation of 2+ state
M. Freer { et al.},
%,H. Fujita, { et al.},
%Z. Buthelezi, J. Carter, R. W. Fearick, S. V. F$\ddot{\rm o}$rtsch, R. Neveling, S. M. Perez, P. Papka, F. D. Smit, J. A. Swartz,  I. Usman,
Phys. Rev. C {\bf 80}, 041303(R) (2009).
%12C second 0+ band states 
\bibitem {Zimmerman2011} 
W. R. Zimmerman{ et al.},
%, N. E. Destefano, M. Freer, M. Gai,  F. D. Smit,
Phys. Rev. C {\bf 84}, 027304 (2011).
\bibitem {Zimmerman2013}% Unambiguous Identification of the Second  State in 12C and the Structure of the Hoyle State
W. R. Zimmerman { et al.},
%, M. W. Ahmed, B. Bromberger, S. C. Stave, A. Breskin, V. Dangendorf, Th. Delbar, M. Gai, S. S. Henshaw, J. M. Mueller { et al.},
%, C. Sun, K. Tittelmeier, H. R. Weller,  Y. K. Wu,
Phys. Rev. Lett. {\bf 110}, 152502 (2013).
\bibitem {Freer2011} % observation of 4+ state
M. Freer { et al.},
%S. Almaraz-Calderon, A. Aprahamian, N. I. Ashwood, M. Barr, B. Bucher, P. Copp, M. Couder, N. Curtis, X. Fang { et al.},
%, F. Jung, S. Lesher, W. Lu, J. D. Malcolm, A. Roberts, W. P. Tan, C. Wheldon,  V. A. Ziman,
Phys. Rev. C {\bf 83}, 034314 (2011).
\bibitem {Itoh2011A} 
M. Itoh{ et al.},
%,H. Akimune, M. Fujiwara, U. Garg, H. Hashimoto, T. Kawabata, K. Kawase, S. Kishi, T. Murakami, K. Nakanishi { et al.},
%, Y. Nakatsugawa, B.K. Nayak, S. Okumura, H. Sakaguchi, H. Takeda, S. Terashima, M. Uchida, Y. Yasuda, M. Yosoi, J. Zenihiro,
 Nucl. Phys.   A {\bf 738}, 268 (2004).
 \bibitem {Itoh2011B} 
M. Itoh { et al.},
%H. Akimune, M. Fujiwara, U. Garg, N. Hashimoto, T. Kawabata, K. Kawase, S. Kishi, T. Murakami, K. Nakanishi  { et al.},
%, Y. Nakatsugawa, B. K. Nayak, S. Okumura, H. Sakaguchi, H. Takeda, S. Terashima, M. Uchida, Y. Yasuda, M. Yosoi,  J. Zenihiro,
 Phys. Rev. C {\bf 84}, 054308 (2011).
\bibitem {Itoh2013} 
M. Itoh { et al.},
%,H. Akimune, M. Fujiwara, U. Garg,  T. Kawabata, K. Kawase,  T. Murakami, K. Nakanishi, Y. Nakatsugawa,  H. Sakaguchi  { et al.},
% S. Terashima, M. Uchida, Y. Yasuda, M. Yosoi, J. Zenihiro,
J. Phys. Conf. Ser. {\bf 436}, 012006 (2013).
% 16O BEC
\bibitem{Funaki2008C}
Y.~Funaki { et al.},
%, T.~Yamada, H.~Horiuchi, G.~R\"{o}pke, P.~Schuck,  A.~Tohsaki,
Phys.~Rev.~Lett. {\bf 101}, 082502 (2008).
\bibitem {Takahashi2020}%16O 4 alpha
J.~Takahashi, Y.~Yamanaka,  S.~Ohkubo,
 Prog. Theor. Exp. Phys. {\bf 2020}, 093D03 (2020).
%Prog. Theor. Exp. Phys. 093D03 (2020).
\bibitem{Ohkubo2010}
S.~Ohkubo,  Y.~Hirabayashi,
Phys.~Lett.~B {\bf 684}, 127 (2010).
\bibitem {Funaki2018} %16O container
Y.~Funaki,
Phys.~Rev.~C {\bf 97}, 021304(R) (2018).
%16O 
\bibitem {Chevallier1967}
P. Chevallier, F. Scheibling, G. Goldring, I. Plesser,  M. W. Sachs,
Phys. Rev. {\bf 160}, 827 (1967).
\bibitem{Freer1995} %16O 4 alpha search 12C(16O,4alpha)
M.~Freer { et al.},
%, N. M. Clarke, N. Curtis, B. R. Fulton, S. J. Hall, M. J. Leddy, J. S. Pople, G. Tungate, R. P. Ward, P. M. Simmons  { et al.},
%W. D. M. Rae, S. P. G. Chappell, S. P. Fox, C. D. Jones, D. L. Watson, G. J. Gyapong, S. M. Singer, W. N. Catford,  P. H. Regan,
Phys.~Rev.~C {\bf 51}, 1682 (1995).
\bibitem{Freer2004}% 8Be+8Be decay of 16O
M.~Freer { et al.},
%, M. P. Nicoli, S. M. Singer, C. A. Bremner, S. P. G. Chappell, W. D. M. Rae, I. Boztosun, B. R. Fulton, D. L. Watson, B. J. Greenhalgh  { et al.},
%G. K. Dillon, R. L. Cowin,  D. C. Weisser,
Phys.~Rev.~C {\bf 70}, 064311 (2004).
\bibitem{Freer2005} %alpha particle states in 16O and 20Ne short contribution
M.~Freer, % for charrisa collaboration
J.~Phys. G {\bf 31}, S1795 (2005).
\bibitem{Itoh2014} %16O 4 alpha search
M.~Itoh { et al.},
%H. Akimune, M. Fujiwara, U. Garg, N. Hashimoto, T. Kawabata, K. Kawase, S. Kishi, T. Murakami, K. Nakanishi { et al.},
%, Y. Nakatsugawa, B. K. Nayak,  H. Sakaguchi,  S. Terashima, M. Uchida, Y. Yasuda, M. Yosoi,  J. Zenihiro,
J.~Phys.~Conf.~Ser. {\bf 569}, 012009 (2014).
\bibitem{Curtis2016} %16O 4 alpha search
N.~Curtis { et al.},
%,S. Almaraz-Calderon, A. Aprahamian, N. I. Ashwood, M. Barr, B. Bucher, P. Copp, M. Couder, X. Fang, M. Freer { et al.},
%, G. Goldring, F. Jung, S. R. Lesher, W. Lu, J. D. Malcolm, A. Roberts, W. P. Tan, C. Wheldon,  V. A. Ziman,
Phys.~Rev.~C {\bf 94}, 034313 (2016).
\bibitem {Barbui2018} %16O 15.1 MeV 
M. Barbui { et al.},
%, K. Hagel, J. Gauthier, S. Wuensche, R. Wada, V. Z. Goldberg, R. T. deSouza, S. Hudan, D. Fang, X. G. Cao  { et al.},
%,  J. B. Natowitz,
Phys.~Rev.~C {\bf 98}, 044601 (2018).

\bibitem{Landau1941}
L. D. Landau,
J. Phys. (USSR) {\bf 5}, 71 (1941).
\bibitem{Landau1947}
L. D. Landau,
J. Phys. (USSR) {\bf 11}, 91 (1947).

\bibitem{Feynman1953A}
R. P. Feynman,  Phys. Rev. {\bf 91}, 1291 (1953).
\bibitem{Feynman1953B} 
 R. P. Feynman,   Phys. Rev. {\bf 91}, 1301 (1953).
 \bibitem{Feynman1954}
R. P. Feynman, Phys. Rev. {\bf 94}, 262 (1954).
\bibitem{Feynman1957} 
 R. P. Feynman,  Rev. Mod. Phys. {\bf 29}, 205 (1957).
 
\bibitem{Josephson1966}
B.~D. Josephson,
Phys. Lett. {\bf 21}, 608 (1966).
\bibitem{Broglia2022}%Transient Joule- and (ac) Josephson-like photon emission in one- 
R. A. Broglia { et al.},
%, F. Barranco, G. Potel,  E. Vigezzi,
Phys. Rev. C {\bf 105}, L061602 (2022).

\bibitem {Nakamura2014}
Y.~Nakamura, J.~Takahashi,  Y.~Yamanaka,
Phys.~Rev.~A {\bf 89},  013613 (2014). 
\bibitem{Griffin1993}
A. Griffin, {\it Excitations in a Bose Condensed Fluid} (Cambridge Studies in Low Temperature Physics, Vol. 4),
edited by A. M. Goldman, P. V. E. McClintock,  M. Springford (Cambridge University Press, Cambridge, 1993).
\bibitem{Glyde1995}
H. R. Glyde, {\it Excitations in Liquid and Solid Helium}  (Oxford Series on Neutron Scattering in Condensed
Matter, Vol. 9), edited by S. W. Lovesey, E. W. J. Mitchell (Oxford University Press, Oxford, 1995). 

\bibitem{Cowley1971} %neutron scattering Roton Phonon
R. A. Cowley, A. D. B. Woods,
Can. J. Phys. Cond. Mat. {\bf 49}, 177 (1971). 
\bibitem{Woods1973} %neutron scattering Roton Phonon
A. D. B. Woods, R. A. Cowley,
Rep. Prog. Phys. {\bf 36}, 1135 (1973).
\bibitem{Andersen1994} % Collective excitations in liquid 4He: I. Experiment and presentation of data
K. H. Andersen {et al.}, 
%W. G. Stirling, R. Scherm, A. Stunault, B. F\r{a}k, H. Godfrin,  A. J. Dianoux,
J. Phys. Cond. Mat. {\bf 6}, 821 (1994).
\bibitem{Andersen1994B}%Collective excitations in liquid 4He: II. Analysis and comparison with theory
K. H. Andersen, W. G. Stirling,
J. Phys. Cond. Mat. {\bf 6}, 5805 (1994). 
\bibitem{Gibbs1999}% The collective excitations of normal and superfluid 4He: the dependence on pressure and temperature
M. R. Gibbs  {et al.}, 
%, K. H. Andersen, W. G. Stirling,  H. Schober,
J. Phys. Cond. Mat. {\bf 11}, 603 (1999).
\bibitem{Milner2023}% HeII Controlled Excitation of Rotons in Superfluid Helium with an Optical Centrifuge
A. A. Milner, V. Milner,
Phys. Rev. Lett. {\bf 131}, 166001 (2023).
\bibitem{Palevsky1957} %Excitation of Rotons in Helium II by Cold Neutrons 
H. Palevsky  {et al.}, 
%, K. Otnes, K. E. Larsson, R. Pauli,  R. Stedman,
Phys. Rev. {\bf 108}, 1346 (1957).
\bibitem{Vinen1958} % observation of roton in He II
W. F. Vinen, Nature {\bf 181}, 1524 (1958).
\bibitem{Dell2003}%Rotons in Gaseous Bose-Einstein Condensates Irradiated by a Laser
D. H. J. O'Dell, S. Giovanazzi,   G. Kurizki,
Phys. Rev. Lett. {\bf 90}, 110402 (2003).
\bibitem{Santos2003}%Roton-Maxon Spectrum and Stability of Trapped Dipolar Bose-Einstein Condensates
L. Santos, G. V. Shlyapnikov,  M. Lewenstein,
Phys. Rev. Lett. {\bf 90}, 250403 (2003).
\bibitem{Blakie2012}%Roton spectroscopy in a harmonically trapped dipolar Bose-Einstein condensate
P. B. Blakie, D. Baillie,  R. N. Bisset,
Phys. Rev. A {\bf 86}, 021604(R) (2012).
\bibitem{Bisset2013}%Fingerprinting Rotons in a Dipolar Condensate: Super-Poissonian Peak in the Atom-Number Fluctuations
R. N. Bisset, P. B. Blakie,
Phys. Rev. Lett. {\bf 110}, 265302 (2013).
\bibitem{Schmidt2021}%Roton Excitations in an Oblate Dipolar Quantum Gas
J-N. Schmidt { et al.},
%, J. Hertkorn, M. Guo, F. B\"{o}ttcher, M. Schmidt, K. S. H. Ng, S. D. Graham, T. Langen, M. Zwierlein, T. Pfau,
Phys. Rev. Lett. {\bf 126}, 193002 (2021).

\bibitem{Chomaz2018}%Observation of roton mode population in a dipolar quantum gas
L. Chomaz { et al.},
%, R. M. W. van Bijnen, D. Petter, G. Faraoni, S. Baier, J. H. Becher, M. J. Mark,
%F. W\"{a}chtler, L. Santos,  F. Ferlaino,
Nat. Phys. {\bf 14}, 442 (2018).
\bibitem{Petter2019}%Probing the Roton Excitation Spectrum of a Stable Dipolar Bose Gas
D. Petter { et al.},
%, G. Natale, R. M. W. van Bijnen, A. Patscheider, M. J. Mark, L. Chomaz,  F. Ferlaino,
Phys. Rev. Lett. {\bf 122}, 183401 (2019).
%supersolid
\bibitem{Andreev1969}
A. F. Andreev, I. M. Lifshitz,
Sov. Phys. JETP {\bf 29}, 1107 (1969).
\bibitem{Chester1970} %1969 submit speculations on BEC and crystals
G. V. Chester, Phys. Rev. A {\bf 2}, 256 (1970).
\bibitem{Leggett1970}%1970 Sep submit Is sold Be is a superfluid
A. J. Leggett, Phys. Rev. Lett. {\bf 25}, 1543 (1970).
\bibitem{Matsuda1970}%Off-Diagonal Long-Range Order in Solids 1970.nov submit
H. Matsuda, T. Tsuneto,
Suppl. Prog. Theor. Phys. {\bf 46}, 411 (1970).
%superfluid discovery
\bibitem{Kapitsa1938}%Viscosity of Liquid Helium Below the lambda-Point
P. Kapitza, Nature {\bf 141}, 74 (1938).
\bibitem{Allen1938}%Flow of Liquid Helium II
J. F. Allen, A. D. Misener, Nature {\bf 142}, 643 (1938).
%optical supersolid 
\bibitem{Tanzi2019} %Observation of a dipolar quantum gas with metastable supersolid properties,
L. Tanzi { et al.},
%, E. Lucioni, F. Fam\`{a}, J. Catani, A. Fioretti, C. Gabbanini, R. N. Bisset, L. Santos,  G. Modugno, 
Phys. Rev. Lett. {\bf 122}, 130405 (2019).
\bibitem{Bottcher2019}% Transient Supersolid Properties in an Array of Dipolar Quantum Droplets,
F. B$\ddot{\rm o}$ttcher { et al.},
%, J.-N. Schmidt, M. Wenzel, J. Hertkorn, M. Guo, T. Langen,  T. Pfau, 
Phys. Rev. X {\bf 9}, 011051 (2019).
\bibitem{Chomaz2019} %Long-Lived and transient supersolid behaviors in dipolar quantum 
L. Chomaz { et al.},
%,   D. Petter,   P. Ilzh\"{o}fer,   G. Natale, A. Trautmann, C. Politi, G. Durastante, R. M. W. van Bijnen, A. Patscheider
%M. Sohmen { et al.}, 
% ,%   M. J. Mark,  F. Ferlaino,
Phys. Rev. X {\bf 9}, 021012 (2019).
\bibitem{Biagioni2024}
G. Biagioni { et al.},
%, N. Antolini, B. Donelli, L. Pezz$\grave{e}$, A. Smerzi, M. Fattori, A. Fioretti, C. Gabbanini, M. Inguscio, L. Tanzi,   G. Modugno,
Nature, {\bf  629}, 773 (2024).
\bibitem {Ohkubo2020B}%12C supersolid
S.~Ohkubo, J.~Takahashi,  Y.~Yamanaka,
 Prog. Theor. Exp. Phys. {\bf 2020}, 041D01 (2020).
%Prog. Theor. Exp. Phys. 041D01 (2020).
\bibitem {Ohkubo2022}%40 supersolid
S.~Ohkubo, 
Phys.~Rev.~C {\bf 106}, 034324 (2022).

\bibitem {Adachi2020}
S. Adachi { et al.},
%,  Y. Fujikawa, T. Kawabata, H. Akimune, T. Doi, T. Furuno, T. Harada, K. Inaba, S. Ishida, M. Itoh { et al.},
 %, C. Iwamoto, N. Kobayashi, Y. Maeda, Y. Matsuda, M. Murata, S. Okamoto, A. Sakaue, R. Sekiya, A. Tamii,  M. Tsumura,
Phys. Lett. B {\bf  819}, 136411 (2021).
\bibitem {Swartz2015}% tentative candidate 5 alpha in 20Ne
J. A. Swartz { et al.},
%,  B. A. Brown, P. Papka, F. D. Smit, R. Neveling, E. Z. Buthelezi, S. V. F$\ddot{\rm o}$rtsch, M. Freer, Tz. Kokalova, J. P. Mira { et al.},
%, F. Nemulodi, J. N. Orce, W. A. Richter,  G. F. Steyn,
Phys. Rev. C {\bf 91}, 034317 (2015).
\bibitem{Zhou2023}
B. Zhou { et al.},
%, Y. Funaki, H. Horiuchi, Yu-Gang Ma,  G. R$\ddot{\rm o}$pke,  P. Schuck,  A. Tohsaki,   T. Yamada,
Nat. Com. {\bf 14}, 8206 (2023).
\bibitem {Ali1966}
S. Ali, A. R. Bodmer,
Nucl.~Phys. A {\bf 80}, 99 (1966).
\bibitem {Yamada2004}
T.~Yamada, P.~Schuck, Phys.~Rev.~C {\bf 69}, 024309 (2004).
\bibitem {Kurokawa2005}
C. Kurokawa, K. Kato,
Phys. Rev. C 71, 021301(R) (2005).
\bibitem {Kurokawa2007}
C. Kurokawa, K. Kato,
Nucl. Phys.  A  {\bf 792}, 87 (2007).
\bibitem {Ohtsubo2013}
S.  Ohtsubo, Y. Fukushima, M. Kamimura,  E. Hiyama,
 Prog. Theor. Exp. Phys. {\bf 2013}, 073D02 (2013).
%Prog. Theor.  Exp. Phys.  073D02 (2013).
\bibitem {Ohtsubo2014}
S.  Ohtsubo, Y. Fukushima, M. Kamimura,  E. Hiyama,
J. Phys.  Conf. Ser. {\bf 569}, 012070 (2014). 
\end{thebibliography}
\end{document}